Studying the impact of the class tournament as an assessment method on student achievements in physics courses

**Daniel Dziob[1], Łukasz Kwiatkowski[2], Dagmara Sokołowska[1],**

[1] Smoluchowski Institute of Physics, Jagiellonian University, Krakow, Poland
[2] Department of Econometrics and Operations Research, Cracow University of Economics, Krakow, Poland

ABSTRACT

Testing knowledge is an integral part of summative assessment at schools. It can be performed in many different ways. In this study we propose assessment of physics knowledge by using a tournament approach. Prior to a statistical analysis of the results obtained over a tournament organized in one of Polish high schools, all its specifics are discussed at length, including the types of questions assigned, as well as additional self- and peer-evaluation questionnaires, constituting an integral part of the tournament. The impact of the tournament upon student improvement is examined by confronting the results of a control test with pre-tournament students' achievements reflected in scores earned in former, written tests. We also present some of students' feedback on the idea of tournament as a tool of assessment. Both the analysis of the tournament results and the students' opinions point to at least several benefits of our approach.

## I. INTRODUCTION

Testing knowledge is an integral part of educational assessment, the latter being a process of documenting content knowledge, skills, attitudes and beliefs, usually focused on an individual learner or a learning community as a whole. The most popular distinction in types of assessment is founded upon the difference between formative and summative assessment [1, 2 and references therein; 3-5]. In general, the formative assessment is carried out throughout a unit (course, project), whereas the summative one - at the end of a unit (course, project) [4, 5]. Some authors seem to distinguish between these types of assessment arguing that the summative assessment is "assessment *of* learning", while the formative one is "assessment *for* learning" [6-9].

Focusing on the summative assessment (SA), we can point to three major criteria defining it: i) SA is used to determine whether students have learned what they were expected to learn [5, 9, 10] ; ii) SA is carried out at the end of a specific teaching period, and therefore it is generally of an evaluative nature, rather than diagnostic one [5, 9, 10] ; iii) SA results are often recorded as scores or grades that are then factored into a student permanent academic record [9, 11, 12].

Summative assessment can be performed in many ways [4, 13-15], though written tests are still the most prevalent [16, 17]. However, in many fields a few researchers have come up with an idea of carrying out assessment in some alternative manners [18-20]. These include, in particular, different forms of a written test, extensively described and compared in the literature, such as free- and multiple-response tests [21], a concept test (such as the Test of Understanding Graphs in Kinematics [22], Force Concept Inventory [23] or Brief electricity and magnetism assessment [24]), a constructed-response test [25], an essay test [26] and others. On the other hand, some authors propose to blend formative and summative assessment techniques. According to [27], such a combination, named "formative summative assessment", entails reviewing exams with students so that they get feedback about their comprehension of concepts. Nowadays, we can find different proposals of combining these two types of assessments [28-31], and the boundaries between them become more and more vague. One example of such an approach is "collaborative testing" – an idea of giving students the opportunity for working in groups during an exam [32], at the end of an individual exam [33] or, more often, after the first, but before the second exam taken individually [34-36] (the last two are sometimes named "two-stage exams"). Research has shown that there are many benefits of utilizing collaborative testing as a constructivist learning method. They are described in detail in [37-39] and references therein.

In our study, we use a tournament – a competitive game between groups in the classroom – as a tool for summation. It can be considered as a kind of "collaborative testing", but unlike the forms mentioned above, we first conduct a group exam (distinguishing individual students' marks through their involvement and contribution in the group work), and, secondly, provide a control, individual test (only for the purpose of research, not influencing students' final marks). Following [9], where also the idea of "assessment *as* learning" is introduced (and in which student self-assessment, and, thereby, self-motivation are brought into focus), we design an alternative form of testing knowledge, combining the assessment with learning at the same time. And by learning we mean not only the subject matter itself, but also acquiring and developing other skills, as well as stimulating positive, both intra- and interpersonal dispositions, such as self-motivation, language skills and group work in the form of cooperative learning [40-42].

## II. RESEARCH DESIGN

In this section we provide details on the tournament itself, including its organization, questions assigned and relevant evaluation procedures.

### A. Tournament organization

The tournament was performed in a high school in Wolbrom (a small town of ca. 9 000 inhabitants, in the South of Poland), and it involved 30 students in their final class (K-12). At the time the class had just accomplished a 22-hour course on electricity.

At the beginning of the actual event (lasting for 2.5 lesson hours), the students were divided randomly into 5 groups, by lottery, drawing out lots with names of fairytale heroes upon entering the classroom. After drawing a card with the name of a hero, every student held a seat at one of the five tables, each grouping heroes of one of five fairy tales. Then, actual tournament started. Figure 1 presents the scheme of the entire process. The tournament began with open questions and multiple-choice questions with an increasing level of difficulty, and, therefore, an increasing number of available points (which were the reward for every correct answer). Further, calculus and some practical tasks were assigned. Finally, all groups faced an extra, common task, with the elements of time competition (the winning team was the first one that rang the bell and provided the correct solution to the problem faced by all teams at the same time). At the end, the students were asked to fill in a special self- and peer-assessment questionnaire. After a week, a control test was performed. At each stage, the whole process was monitored by two independent teachers (apart from the major teacher of the class), who were responsible for the verification of verity and integrity of the student evaluation.

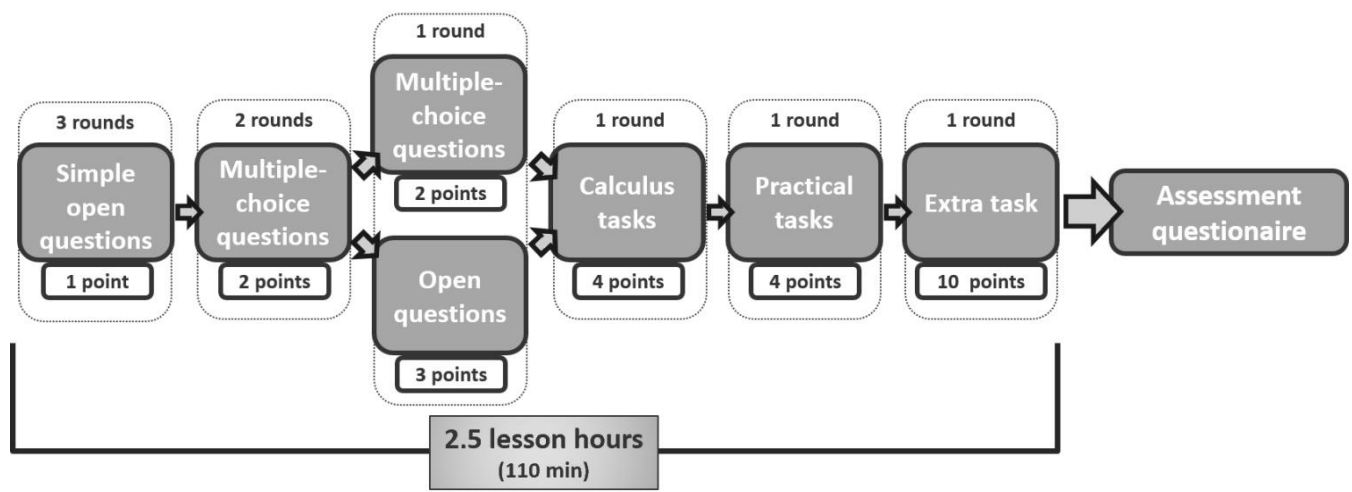

FIG. 1. Tournament testing sequence.

The first two stages were organized in multiple rounds. The first phase comprised three rounds, and the second – two rounds. At the first three stages, in each round the teams attempted a task in consecutive turns. At the third phase the students faced a choice of undertaking either a 3-point open question or a 2-point multiple-choice one. It was intended to introduce some element of decision-making risk, thereby facilitating students' sense of responsibility for the choices made. The following three stages (i.e. the calculus, practical and extra tasks) were single-rounded and at each of them all teams were challenged with their tasks at the same time.

In the first three types of questions students from the currently "active" group were required to choose the number of a question, and then the team had the appointed time (respectively 30 seconds, 1 minute, 2 minutes) to deliver the answer. If they did not succeed or their answer was incorrect, other groups could take over the question and score extra points by ringing the bell and providing the correct answer. Allowing for such a possibility was meant to ensure attention and an active interest of each group in the question currently dealt with by any other team. During this part of the tournament, questions were projected onto the wall screen so as to make it available for all teams at the same time. In the calculus and practical tasks all groups worked simultaneously over different, randomly selected problems, received on sheets of paper. The extra task was the same for all groups, and, again, it was projected on the wall screen so as to make it available to all teams at the same time. The first group which solved the problem won (according to the rule "first-come, first-win"), and scored extra 10 points.

*1. Simple open questions (1 point)*

Open questions, each 1-point worth, were meant "to warm up" the students. The tasks were related to some basic knowledge from the curriculum, requiring the students to provide correct simple formulae, units etc., and also examining their basic context knowledge (see Fig. 2).

*Q4. What is an electric current? What is the conventional direction of a current flow?*
*Q6. What is an electrolyte?*
*Q8. What is the unit of an electric current?*
*Q9. Give three examples of using Joule heat in everyday life.*
*Q12. Suggest a formula describing the relation between the temperature and the resistance for conductors.*

FIG. 2. Examples of simple open questions (1 point).

*2. Multiple-choice questions (2 points)*

Then, two rounds ensued of multiple-choice scientific reasoning questions (each worth 2 points). The students were requested not only to point out the correct answer, e.g. "C", but also to provide a proper explanation of their choice (see Fig. 3).

*Q2. Ah, this heat*

*For all metals, an increase of the temperature causes an increase in resistance. What is the effect of temperature increase?*

*A) an increase in the oscillation speed of both atoms and ions*

*B) an increase in the density of the electron gas*

*C) affecting the binding of/from the metal to the valence*

*D) deterioration of contact between the microcrystals*

(Correct answer: A)

*Q7. Absent-minded electrician*

*When building an electric circuit somebody swapped voltmeter with ammeter. What happens when you turn on the voltage source?*

*A) ammeter burns out*

*B) voltmeter burns out*

*C) both voltmeter and ammeter burn out*

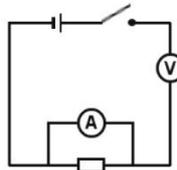

*D) meters do not burn out, and the current decreases to almost zero*

(Correct answer: D)

FIG. 3. Examples of multiple-choice questions (2 points).

### 3. Open questions (3 points)

At the third stage of the tournament, each group faced a choice between a multiple-choice question worth 2 points and an open question for 3 points. The latter was more challenging, requiring broader knowledge and ability of connecting facts, see Fig. 4.

*Q1. The figures below show the relationship between the resistivity and the temperature for three different materials. Which of the graphs corresponds to metal, superconductor, semiconductor? Assign and justify.*

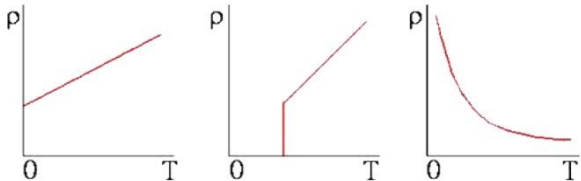

*Q5. Draw a scheme of an electrochemical cell. What determines the voltage obtained from a given cell?*

FIG. 4. Examples of open questions (3 points).

*4. Calculus tasks (4 points)*

After this part, calculus tasks followed. Each group had to choose a different problem (see Fig. 5 for an example) and was given 10 minutes to provide the correct solution. As previously mentioned, this time all groups worked simultaneously. As a result each team could receive maximally 4 points, with a lower score given upon delivery of either an incomplete or partially faulty solution.

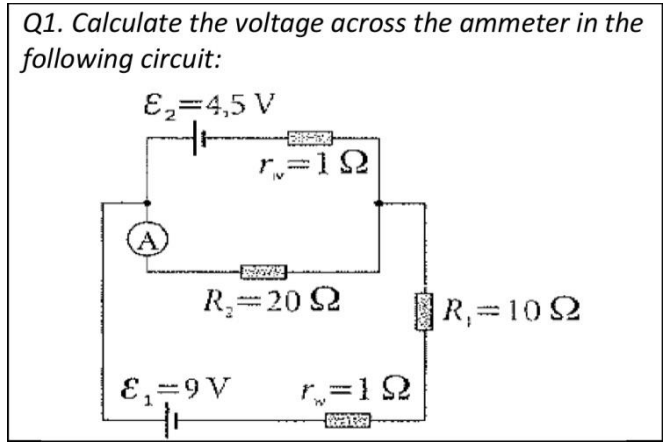

FIG. 5. Example of a calculus task.

*5. Practical tasks (4 points)*

The penultimate challenge was a practical task. Each group had to pick randomly an experimental task on one of the following six themes: galvanization, electrochemical cell, electrolysis of water, Ohm's law, building a circuit according to an assigned scheme, and voltage measurement in a designated point. Each team was requested to build a proper circuit, carry out the experiment and give valid description and explanation of the phenomenon at hand. All necessary equipment in each case, with some redundant materials mixed in, was available on a table. Then, students had to decide by themselves which objects were indispensable to accomplish the task.

*6. Extra task (10 points)*

The final and most demanding task was common to all groups. For all groups at the same time, a slide with the "Monstrous maze of resistors" adopted from [43] was displayed on the wall screen. The first group which found the solution and gave the correct answer received 10 points.

*7. Assessment questionnaires*

After the tournament each student was asked to fill in individually special self- and peer-assessment questionnaires, aimed at evaluating himself/herself and other fellow players from the same group in various aspects. Each of eight questions required allotting a score between 1 and 6. Four of them were related to the "communication skills", whereas the other four were focused on assessing the "subject matter contribution". In the table below we present the self-assessment questionnaire. The peer-assessment questions were designed analogously.

TABLE I. Student self-assessment questionnaire.

| Question | 1-6 scale |
|---|---|
| Were you involved in the work group? | Communication skills |
| Did you communicate adequately in the group? | |
| Did you take part in the discussion on the problem? | |
| Did you take into account the opinion of others? | |
| Did you prepare for the test beforehand? | Subject matter contribution |
| Did you take part in solving problems and tasks? | |
| Did you have sufficient knowledge to solve the issues? | |
| Did you contribute to the final result of the group? | |

## B. Evaluation process

The final note granted to each student consisted of three components:

I.   the group percentage result from the tournament questions (the first six stages) – with a weight of 0.5
II.  the questionnaire-based assessment result for the "subject matter contribution" – with a weight of 0.3
III. the questionnaire-based assessment result for the "communication skills" with a weight of 0.2.

The percentage score for each team was obtained through dividing the number of points accumulated by the group by the maximal number of points possible to obtain. The points scored for answering the questions taken over from other groups were not included in the maximal number of possible points.

The questionnaire-based assessment results were included in the final score according to the authors' own approach presented in the diagram bellow. For each person, the algorithm proceeded as follows:

1) Firstly, the median score was calculated of "subject matter contribution" and, separately, "communication skills" points in the self-assessment results (S).
2) Secondly, the median score was calculated of "subject matter contribution" and, separately, "communication skills" points attributed to the student by all other members of the group (the peer-assessment, P).
3) Then, the "subject matter contribution" and "communication skills" scores were obtained separately according to the rule:
   • If $|S - P| \leq 1$ (a consistent evaluation): take P as the final score
   • else (an inconsistent evaluation): take $P - 0.5$ as the final score

There are three premises behind the above algorithm. Firstly, we choose to represent the "average" (benchmark) score (in both S and P) by a median rather than a mean, for the previous – as opposed to the latter – is robust to extremities. Secondly, the assumed value of "1" as a tolerable discrepancy between S and P still ensuring a consistent evaluation is our arbitrary choice that appears justifiable in view of the 6-point scale employed in the questionnaires. Finally, in the case of an inconsistent evaluation we penalize the P result with an arbitrary value of 0.5. Note that regardless of the precise relation between the S and P assessments, the penalization is always downward, which is intended to reduce a risk of „collusion" among the students, and to stimulate honest and reasonable both self- and peer-assessments (the students had been familiarized with the algorithm prior to the tournament). The final score in the tournament, calculated according to the algorithm above, is henceforth denoted as "TNT".

## III. DATA COLLECTIONS AND ANALYSIS

In this section we provide details about the pre-tournament test and the post-tournament (control) test, to assess students' progress (attributable to the tournament) with respect to their former achievements. To this end, a statistical analysis of relevant results is further performed.

### A. Control test

The control test was prepared in a traditional, written form, and conducted one week after the tournament. The test was unannounced, so the students could not have made any additional efforts to prepare for it. In 60% the test comprised tasks utilized during the tournament, and in the remaining 40% it was based on problems totally new to the students, though similar to the ones given in the tournament. The control test score is expressed in percentage terms, and, henceforth, denoted as "CT".

### B. Former tests

Each student, during the school year and before implementation of the tournament, participated in three tests: on thermodynamics, gravitation and electrostatics. All tests were taken individually. They contained mixed problems, including content knowledge and scientific reasoning tasks, multiple choice, open-response and calculus problems. To measure each student's achievements prior to the tournament, we use the average of his/her results on the three tests. The quantity obtained (expressed in percentage terms) is further referred to as the "former tests score" and denoted as "FT".

### C. Basic statistical analysis

Figure 6 presents each student's three individual scores: on the former tests (FT), the tournament (TNT), and the control test (CT), along with horizontal bars indicating the common (for each group) result gained from the tournament. All scores are provided in percentage terms. Note that the discrepancies between a common result in a group and its members' individual scores stem from the outcomes obtained in the assessment questionnaires. Notice that, incidentally, the final marks assigned to each student within the fifth team were all lower than the common result of ca. 95%. This observation can be explained by the fact that nobody in the group was perceived as a leader, and all the team members were clearly aware of the fact that their final result was the effect of their cooperation (rather than attributable to the knowledge of a single leading person).

It can be easily noticed that the TNT marks were predominantly way above the FT results. What appears far more justifiable, however, is the comparison of the student achievements and skills prior to and after the tournament, reflected in the FT and CT results, respectively. In that regard, however, we still observe a systematic (i.e. for almost all tournament participants) increase in score, with the result hinting at a positive impact of the tournament on the students' improvement.

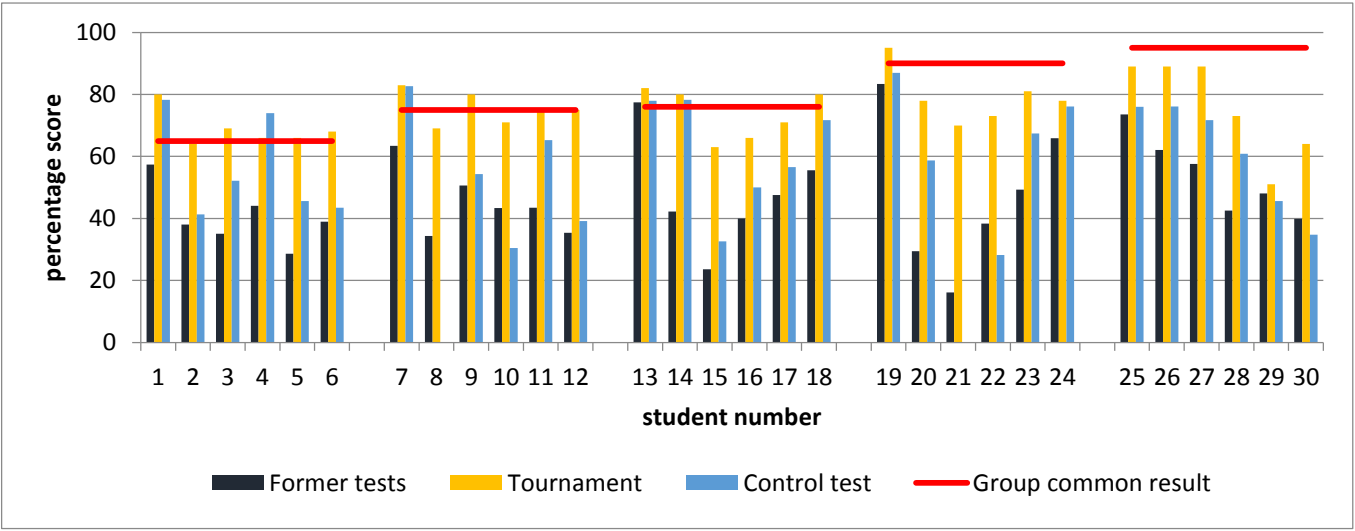

FIG. 6. Student scores. For each student the three vertical bars represent (starting with the leftmost one): the average score in former tests, the final score obtained in the tournament, and the mark gained in the post-tournament, control test. The horizontal lines represent the common result scored by each group in the tournament (based on the first six stages).

In what follows, to explore the results in more detail, we conduct statistical analysis. As far as the sample size is concerned, since two students (no. 8 and 21; see Fig. 6) were absent from the control test, we exclude them from further considerations, and carry out the necessary calculations based on the sample of n = 28 students.

In Table II and Figure 7 we present basic descriptive statistics and empirical distributions (histograms, with normality tested by the Lilliefors and the Shapiro-Wilk tests) for several variables, including the FT, TNT and CT scores, as well as the differences: TNT – FT and CT – FT (the latter measuring the "absolute" gain in student content knowledge). Moreover, we also examine a modified gain factor (MGF), which is our adaptation of the normalized gain (or the g-factor) [44], originally proposed in [45]. The MGF measure is meant to relate the "absolute" gain in a student CT score to the points missed on FT, and is therefore calculated according to the formula:

$$MGF = \frac{CT - FT}{100 - FT}.$$

TABLE II. Basic statistics of the student results, including: the average score in the former tests (FT), the final score in the tournament (TNT), the result in the post-tournament, control test (CT), differences between TNT and FT (TNT − FT), as well as CT and FT (CT − FT). The last row contains statistics for the modified gain factor (MGF).

| Variable | Characteristics | | | | | | |
|---|---|---|---|---|---|---|---|
| | Mean | 95%-confidence interval for mean | Median | Lower quartile | Upper quartile | Interquartile range | Standard deviation |
| FT | 48.39 | (42.64; 54.14) | 43.79 | 38.65 | 57.48 | 18.83 | 14.83 |
| TNT | 74.96 | (71.17; 78.76) | 75.00 | 67.00 | 80.50 | 13.50 | 9.78 |
| CT | 59.16 | (52.29; 66.04) | 59.78 | 44.57 | 76.09 | 31.52 | 17.73 |
| TNT − FT | 26.57 | (22.49; 30.65) | 27.28 | 22.27 | 32.82 | 10.55 | 10.52 |
| CT − FT | 10.77 | (6.22; 15.32) | 10.11 | 3.44 | 18.22 | 14.78 | 11.73 |
| MGF | 0.22 | (0.13; 0.3) | 0.23 | 0.07 | 0.37 | 0.30 | 0.22 |

From Table II it can be inferred that the students scored, on average, ca. 48.4% upon the former tests, with the standard deviation hovering around 14.8 percentage points (henceforth, pp). Half of the students recorded the FT result below ca. 43.8%, whereas the other half − above that number. (The means and medians differ on account of positive skewness of the empirical distribution; see Fig. 7(a)). On the other hand, results obtained during the tournament are distinctive on two counts. Firstly, the average TNT score is much higher as compared to FT, with the difference being statistically significant at any typical $\alpha$ level (i.e. $\alpha \in \{0.01, 0.05, 0.1\}$; see Tab. III). Arguably, the difference can be attributed to the team work and cooperation among the students. (The latter argument appears particularly valid for the last group, for the reasons already explained above). Secondly, the TNT scores are more concentrated (as compared with FT) around the mean, with a drop in standard deviation of ca. 5 pp. Moreover, the TNT distribution is far more symmetrical than its FT counterpart (see Fig. 7(b)), thereby closing the gap between the mean and median (both equal around 75%; see Tab. II). In general, the TNT scores are more regularly, symmetrically distributed and strongly shifted rightwards as compared to the FT results. (Note, however, that for all but one the analyzed variables, with CT being the exception, despite more or less conspicuous irregularities such as skewness and multimodality, the null hypothesis of normality is not rejected, which, admittedly, is largely due to the low sample size. Still, as implied by the corresponding p-values, the TNT distribution is far closer to normal than actually any of the others; see Fig. 7). In practical terms, it can be inferred that a tournament, as a form of assessment, yields higher, less dispersed and more normally distributed outcomes than traditional forms of a test (see Fig. 7(a-c)).

TABLE III. Testing positive means for: the difference between the scores earned in the tournament and the former tests (TNT − FT), the difference between the results gained in the control test and the former tests (CT − FT), and the modified gain factor (MGF). In the second column values of the Student-t statistics are displayed for testing a positive mean. The last column presents corresponding p-values.

| Characteristics | Test statistics | p-value |
|---|---|---|
| TNT − FT | 13.36 | $1.02 \times 10^{-13}$ |
| CT − FT | 4.86 | $2.20 \times 10^{-5}$ |
| MGF | 5.30 | $6.80 \times 10^{-6}$ |

Moving on to the CT results, it appears, interestingly, that these are somehow less regular than FT, on two counts. Firstly, the CT distribution has a higher dispersion, as implied by both standard deviation and, in particular, interquartile range (see Tab. II). Secondly, as long as the FT distribution features only a single mode (somewhere between 40 and 50%), the CT histogram exhibits a pronounced bimodality. Apparently, the two CT modes correspond with the ones present in the FT and the TNT distributions, with the global CT mode (between 70 and 80%) coinciding with the TNT one, and the second, a local one (between 40 and 50%) – with the FT mode. In statistical terms, one could argue that the CT distribution is a mixture of the FT and TNT distributions. Practically speaking, it could be inferred that the CT scores are formed as a confluence of student prior physical expertise (measured by FT) and the knowledge and skills acquired during the tournament.

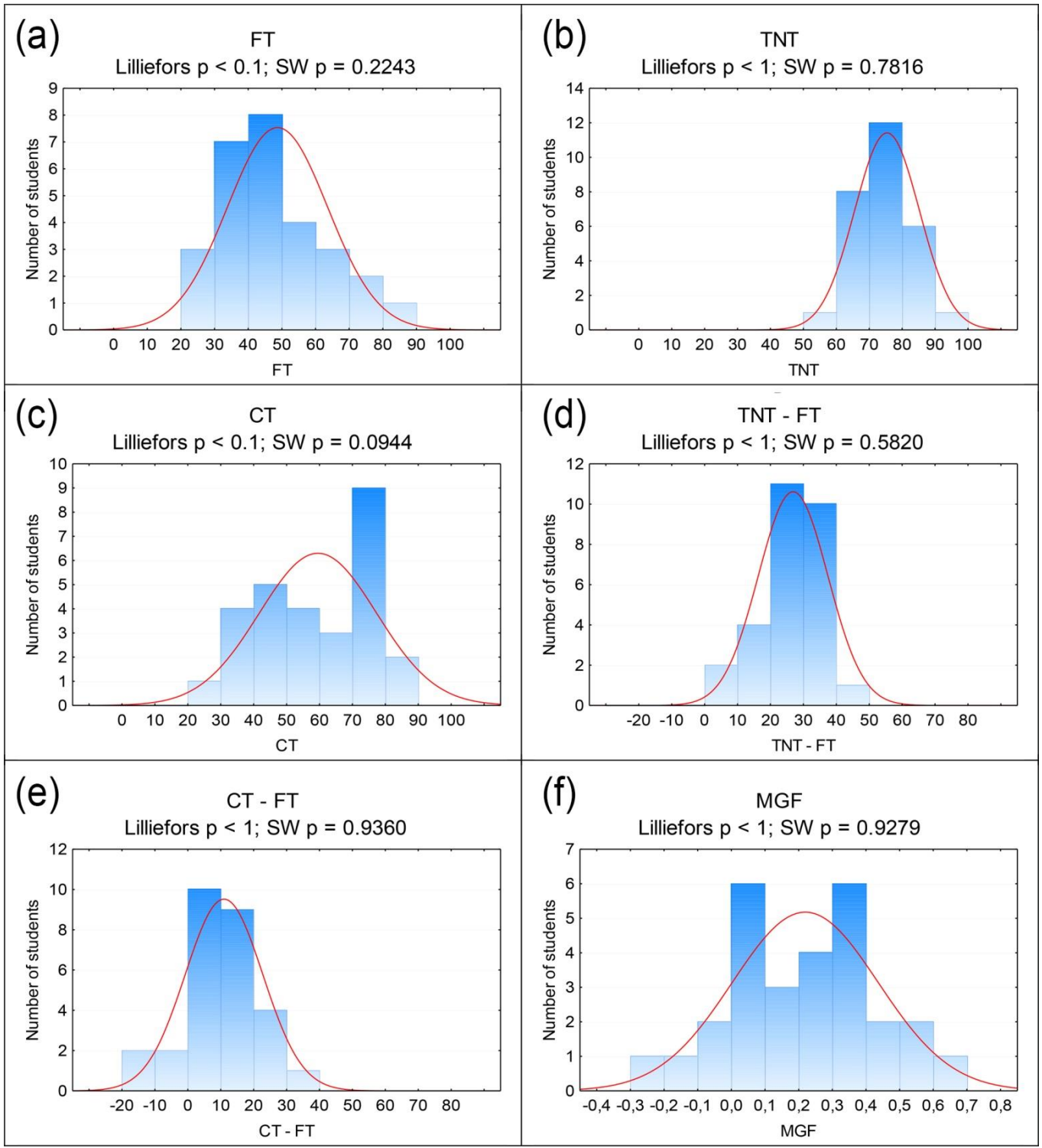

FIG. 7. Histograms of the students' results: (a) the average score in the former tests (FT), (b) the final score in the tournament (TNT), (c) the result in the post-tournament, control test (CT), (d) differences between TNT and FT (TNT − FT), (e) as well as CT and FT (CT − FT). Panel (f) displays the histogram for the modified gain factor (MGF). In each case the normal density is fitted (solid line), accompanied by p-values for testing normality through the Lilliefors and the Shapiro-Wilk tests (denoted as "Lilliefors p" and "SW p", respectively).

Finally, we proceed with the analysis of the results scored in the control test in relation to student content knowledge and skills prior to the tournament (FT results). The average difference between the CT and FT scores totals ca. 10.8 pp (see Tab. II), and it is statistically significant, regardless of the α level (see Tab. III). (Note, however, that four out of 28 students scored lower in CT than in FT, so negative increments were also reported). Improvement of the student performance is also indicated by the results obtained for the modified gain factor. A test of positive MGF mean indicated it is significantly positive at any typical α level (see Tab. III). Note, however, that the MGF histogram exhibits two pronounced and equivalent modes, which may question the use of the mean as a measure of central part of the distribution. Nevertheless, both modes are positive. Furthermore, almost 86% of the probability mass in the histogram is localized to the right of zero, which implies that a learner positive gain was reported for a predominant number of students (i.e. 24 out of 28; see Fig. 7(f)).

<div align="center">D. Correlation and regression analysis</div>

Below, the analysis of correlations between selected pairs of the considered variables is performed. Figure 8 displays relevant scatter plots, along with fitted linear regressions, 95%-confidence bands, linear correlation coefficients (r), p-values for testing their significance, and, at last, the square of correlation coefficients ($r^2$), which coincide with determination coefficients in the fitted regressions. Based on Figure 8, the following general conclusions can be formulated:

1) A positive and statistically significant correlation between FT and TNT implies that students who performed better prior to the tournament, also scored higher in the tournament (see Fig. 8(a)). It is worth underlining that the value of correlation coefficient (r = 0.7059) is negatively affected by the single outlying score equal 51 (obtained by student no. 29), exclusion of which raises the coefficient value to r = 0.8021.

2) Similarly, the TNT and the CT results are positively and statistically significantly interrelated, indicating that better (likewise, worse) performance in the control test coincides with a higher (respectively, lower) score in the tournament (see Fig. 8(b)).

3) As expected on the basis of the two above observations, there also occurs a positive and significant relation between the CT and FT scores, indicating that high (likewise, low) notes in the control test were mostly obtained by those who already performed high (low, respectively) on the former tests (see Fig. 8(c)).

4) Some slight (r = 0.1259), yet statistically insignificant correlation is observed between MGF and FT, hinting at no dependence of a student gain upon his/her previous performance (see Fig. 8(d)).

5) On the other hand, it appears that student improvement (as measured by MGF) is significantly and positively influenced by the tournament performance, though the correlation coefficient is only ca. 0.38 (see Fig. 8(e)). The result suggests that the learner gain is generally higher in the case of those who scored higher on TNT.

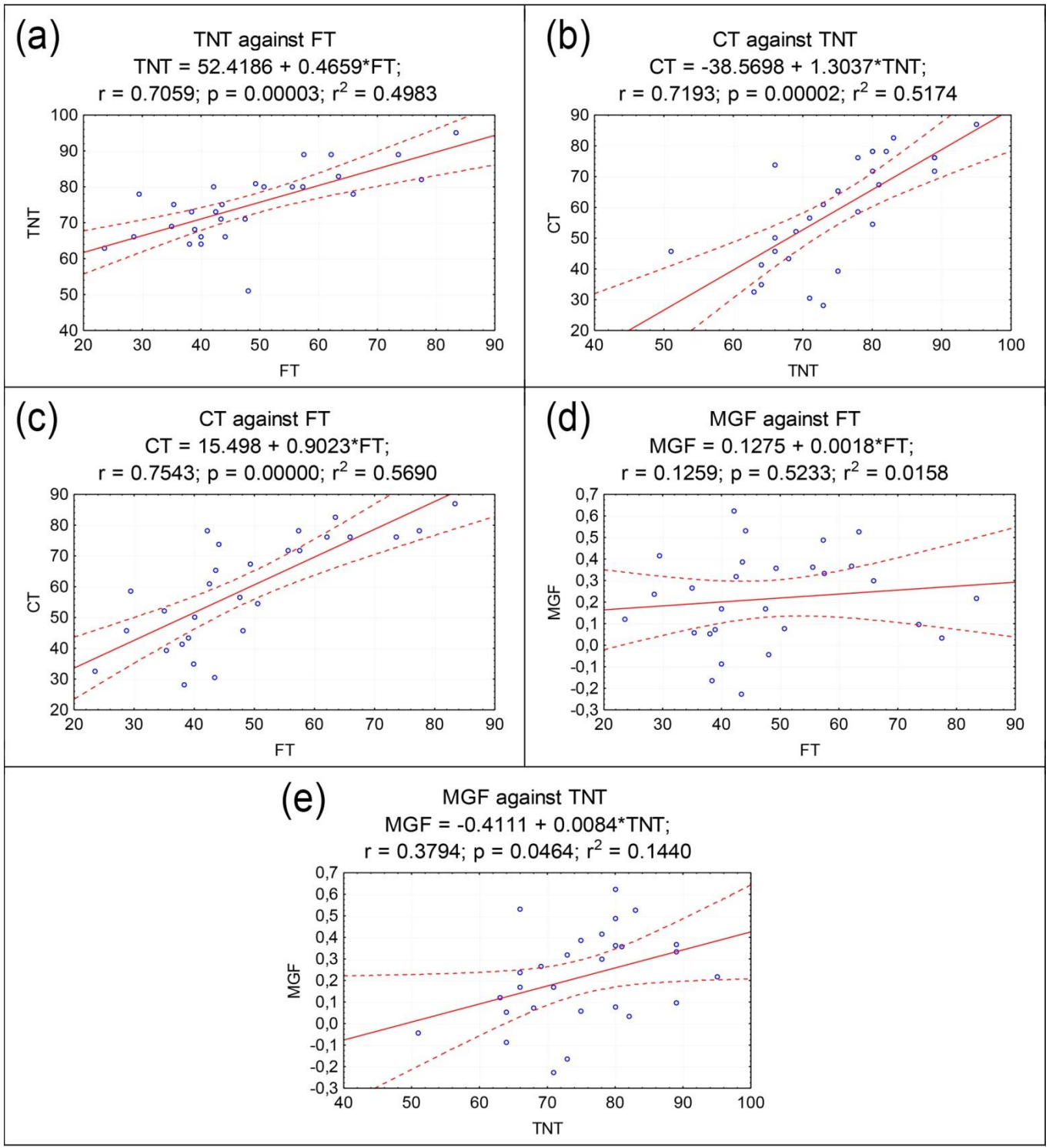

FIG. 8. Scatter plots for selected pairs of the students' results. Apart from data points, in each plot a linear regression is fitted and the 95%-confidence band is marked. Below each regression equation are provided: Pearson's linear correlation coefficient (r), p-value for testing a non-zero correlation coefficient, and determination coefficient ($r^2$).

The inferences formulated in items no. 1-3 point to an intuitive relation according to which the better a student has fared so far, the higher his/her performance in the tournament, and, eventually, in the control test. Further, result no. 4 implies generally that the student gain, arguably attributable to the tournament, hardly depends on his/her former achievements. In broad terms, it would follow that the tournament provided equal opportunities of improvement to all students. Nevertheless, conclusion no. 5 would still indicate that those who performed better in the tournament (as the effect of their active involvement in cooperative work), actually improved slightly (yet significantly) more than the others.

The results presented above provided us an incentive to build two simple bivariate linear regression models in order to jointly evaluate the impact of the former tests and the tournament results on the control test score and the modified gain factor. The two models take the following form:

$$Y = \beta_0 + \beta_1 FT + \beta_2 TNT + \varepsilon,$$

with $\varepsilon$ denoting normally distributed random errors, and $Y$ representing the dependent variable (i.e. either CT or MGF). In Tables IV and V the following estimation results are presented: determination coefficient ($R^2$), point estimates, standard errors, p-values against the alternative of a non-zero coefficient (i.e. $H_1$: *coefficient* $\neq 0$), p-values against the alternative of either a positive or negative coefficient (i.e. $H_1$: *coefficient* $> 0$, or $H_1$: *coefficient* $< 0$), depending on the sign of the point estimate. (Though not reported in the paper, the Lillierfors and Shapiro-Wilk tests do not reject the normality of residuals in any of the regression models considered below, therefore validating testing the regression coefficients by means of a standard Student's t-test).

As regards regressing CT against FT and TNT, it appears that both regressors positively influence the CT score. More specifically, if a student scored higher in FT (likewise, TNT) by 1 pp, then he/she would score also higher in CT by ca. 0.59 pp (0.68, respectively). The results are (positively) significant at $\alpha$ equal 0.01 and 0.05, correspondingly. With respect to the determination coefficient, we note that about 64% of the control test results is explained by the former tests and the tournament performance.

TABLE IV. Regression results for CT. Asterisks indicate statistical significance (a non-zero coefficient): *** for $\alpha$ = 0.01, ** for $\alpha$ = 0.05, and * for $\alpha$ = 0.1.

|  | Dependent variable: CT | | |
| --- | --- | --- | --- |
| Regressor | Intercept | FT | TNT |
| Point estimate | -19.8827 | 0.5878*** | 0.6750** |
| Standard error | 17.6894 | 0.2031 | 0.3077 |
| p-value against a non-zero coef. | 0.2717 | 0.0078 | 0.0378 |
| p-value against a positive/negative coef. | 0.1358 | 0.0039 | 0.0189 |
| Determination coefficient ($R^2$) | | 0.6386 | |

With respect to the regression for MGF, we note that as long as the student gain depends positively on the tournament performance (at $\alpha$ = 0.05), it is not determined by FT (see Tab. V). As already mentioned above, it would follow that the student improvement, arguably attributable to the tournament, does not depend on their former achievements, and, in broad terms, that the tournament provided equal opportunities of improvement to all students. Note, however, that only about 18% of the modified gain factor can be explained by the former achievements and the tournament performance.

TABLE V. Regression results for MGF. Asterisks indicate statistical significance (a non-zero coefficient): *** for $\alpha$ = 0.01, ** for $\alpha$ = 0.05, and * for $\alpha$ = 0.1.

| Regressor | Dependent variable: MGF | | |
|---|---|---|---|
| | Intercept | FT | TNT |
| Point estimate | -0.5421 | -0.0041 | 0.0128** |
| Standard error | 0.3234 | 0.0037 | 0.0056 |
| p-value against a non-zero coef. | 0.1062 | 0.2778 | 0.0320 |
| p-value against a positive/negative coef. | 0.0531 | 0.1389 | 0.0160 |
| Determination coefficient ($R^2$) | | 0.1841 | |

## IV. DISCUSSION

### A. Social benefits

After the control test, and before getting informed about their final marks, the students were asked to express anonymously their opinions about a tournament as a tool of assessment. Some examples of the comments are cited below:

Student A:
*I think that this form of a test is good, because we can share our knowledge with others and vice versa, helping each other. We can memorize more and learn new things.*

Student D:
*This is a better form of consolidation and verification of our knowledge and skills.*

Student E:
*This is a good idea, because it was performed in the form of a game. A student can show what he or she knows without being stressed.*

Student K:
*Fabulous! We can integrate, everybody who had any idea but wasn't sure about it had an opportunity to consult/discuss it with other members of the group.*

Student O:
*I really liked explanation of each answer given afterwards - in this way it was possible to understand more.*

Student W:
*Everybody wanted to receive a good note and knew that there is "collective responsibility" and tried to do his/her best.*

The opinions delivered above lead us to a short discussion about students' social benefits arising from participation in the tournament. In what follows, we relate these with major findings presented in the literature on collaborative testing.

The tournament was organized in the form of a team game, but with elements of rivalry. In this way it can be perceived as a form of activity in which group work skills, desirable in some academic areas and also by employers, are naturally activated, playing crucial role in accomplishing tasks [33, 37, 46-48]. Simultaneously, the tournament induced far less test anxiety (as compared with traditional, individually taken written test) by giving students a sense of being supported by the other team members [33, 37, 48, 49]. Working together may improve communication skills as well. Students learn to listen to each other, share information, and respond to ideas proposed in discussions, which stimulate knowledge assimilation [40, Ch. 3]. What is worth noting is that vocabulary and concepts used in group and class discussions may provide retrieval cues that help students recall relevant information. Moreover, the requirement of providing not only the answer to a question, but also the explanation for it, necessitated that the students should be able to understand and present their lines of reasoning and reconsider them, if needed. Therefore, a tournament may also yield an improvement in student ability of critical thinking [33, 37, 47]. Finally, an active involvement in the self- and peer-assessment process may enhance student confidence and adequate self-esteem [50]. Taking all the above into consideration, cooperative testing of knowledge may become a significant part of a learning process.

## B. Academic benefits

The main purpose of this research was to examine the impact of taking an exam in the form of a tournament on student knowledge. Firstly, a statistically significant increase is observed in students' achievements in the tournament as compared to their former tests results (the average difference amounted to ca. 26 pp, in favor of the tournament scores, being positively significant at any typical $\alpha$ level). Secondly, we also find evidence for improvement of student content knowledge and problem solving skills, as indicated by the results of a control test taken by the students a week after the tournament (the average difference between marks in the control test and former tests scored ca. 11 pp; the mean of modified gain factor totaled 0.22; both results are positively significant at any typical $\alpha$ level). Our findings remain in accordance with much research on positive impact of collaborative testing. Studies presented in [33, 37, 51, 52], focused on the effects of taking exams in a collaborative way for numerous groups with various numbers of students and of different subject/specialization, indicated higher students' achievements as compared with traditional ways of individual testing of knowledge. Moreover, in [51] it was found that collaborative exam scores were also higher than the ones earned in individually taken exams during which students were allowed to use course textbooks and their notes. Further, some researchers show that students' performance also improved in a longer perspective, as indicated by post-tests taken some time after the collaborative exam [36, 53, 54]. Notice that in our research we established a positive and statistically significant impact of participation in the tournament on students' achievements in the control test.

Finally, in the context of the tournament organization, let us emphasize that the event was not preceded by any traditional, individually taken test on the subject matter (i.e. electricity), though, conceivably, it would be worth contrasting the control test results with the ones obtained in a typical pre-test on the same content. In our approach we followed conclusions formulated in [55], who suggests that the learning gain due to taking a collaborative final exam might be higher if the students had no previous individual encounter with relevant tasks. In the cited paper it was found that in the post-test the students scored higher on new problems (i.e. the ones that had not been used in the pre-test) than on the questions they had already been given previously. A possible logic behind this observation is that the lines of reasoning followed by a student during an individually taken exam tend to persist afterwards, therefore hindering acquiring new ways of thinking and solving the problem, even after participating in a collaborative activity. It would follow then that, as claimed in [55], "it might be preferable to collaborate without first deciding on questions individually." Taking this as well as our findings into account, we infer that a class tournament is a well-justifiable and effective learning activity, in which the three approaches to assessment (i.e. *of*, *for* and *as* learning [9]) merge together.

## V. CONCLUDING REMARKS

In this paper results after conducting a knowledge test as a tournament were presented. Based on students' results and opinions we can come up with the following conclusions:

I. Scores obtained during a tournament were higher than in traditionally performed tests.
II. For most learners their results got in an individually written control test (taken a week after the intervention), were higher than their average performance beforehand.
III. The alternative method of testing analyzed in our paper appears to provide equal opportunities of improvement both for low- and high-performers.
IV. Students appreciated the method very much because they could help each other in solving problems in a more cooperative, less stressful way.

In general, there are several advantages of such a form of examination that can be listed, including: supporting weaker students by collaboration with others, setting a framework of cooperative-learning among students, development of group-work skills, stress-free testing, and, in addition to these, integration of the class. Nevertheless, what may appear as some disadvantages of this alternative form of assessment are organizational difficulties. These, however, concern predominantly the teacher rather than the students, for both preparing and conducting a tournament are by far more conceptually demanding and therefore time-consuming than in the case of more traditional testing method. Moreover, we believe that a "full-scale" event itself would usually require reservation of more than a single lesson hour, possibly lasting two or three times as much (recall that the tournament discussed in this study lasted for 2.5 lesson hours). Still, however, it is our stance that the benefits of the approach outweigh additional burden placed on the teacher.

## ACKNOWLEDGEMENTS

Łukasz Kwiatkowski would like to acknowledge financial support by the Foundation for Polish Science (within the START 2014 program).

REFERENCES

[1] P. Black and D. Wiliam, "Assessment and classroom learning", Assessment in Education: Principles, Policy & Practice, 5, 7 (1998)

 [2] D. Wiliam and P. Black, "Meanings and Consequences: a basis for distinguishing formative and summative functions of assessment?" British Educational Research Journal, 22, 537 (1996)

[3] C. Garriso and M. Ehringhaus, (2007), Formative and summative assessments in the classroom. Available online at:
http://www.amle.org/Publications/WebExclusive/Assessment/tabid/1120/Default.aspx

[4] J. Mctighe and K. O'connor, "Seven practices for effective learning", Educational leadership 63, 10 (2005)

[5] H. Wynne and J. Mary, "Assessment and Learning: differences and relationships between formative and summative assessment", Assessment in Education: Principles, Policy & Practice, 4, 365 (1997)

[6]  M. Taras, "Assessment: Summative and Formative: Some Theoretical Reflections",  British Journal of Educational Studies, 53, 466 (2005)

[7] P. Black, C. Harrison, C. Lee, B. Marshall and D. Wiliam, "Working Inside the Black Box: Assessment for Learning in the Classroom", Phi Delta Kappan, 86, 8 (2004)

[8] J. W. Looney, "Integrating Formative and Summative Assessment: Progress Toward a Seamless System?", OECD Education Working Papers, 58, 5 (2011)

[9] L. Earl, (2003) *Assessment as Learning: Using Classroom Assessment to Maximise Student Learning*, (Thousand Oaks, CA, Corwin Press, 2003), chapter 3

[10] H. Torrance and J. Pryor, *Investigating Formative Assessment. Teaching, Learning and Assessment in the classroom*, (Buckingham, Open University Press, 1998)

[11] J. Biggs, "Assessment and classroom learning: A role for summative assessment?" Assessment in Education*, 5, 103; (1998)

[12] B. S. Bloom, J. T. Hastings and G. F. Madaus, *Handbook on the formative and summative evaluation of student learning* (New York: McGraw-Hill, 1971)

[13] P. Black , C. Harrison, J. Hodgen , B. Marshall and N. Serret, "Can teachers' summative assessments produce dependable results and also enhance classroom learning?", Assessment in Education: Principles, Policy & Practice, 18, 451 (2011)

[14] R. W. Tyler, R. M. Gagne and M. Scriven, "The methodology of evaluation", Perspectives of curriculum evaluation, Chicago, 1, 39 (1967)

[15] P. Black , C. Harrison, J. Hodgen , B. Marshall and N. Serret, "Validity in teachers' summative assessments", Assessment in Education: Principles, Policy & Practice, 17, 215 (2010)

[16] S.Vercelatti, M. Michelini, L. Santi, D. Sokolowska and G. Brzezinka, "Investigating MST curriculum experienced by eleven-year-old Polish and Italian pupils " *in E-Book Proceedings of the ESERA 2013 Conference: Science Education Research For Evidence-based Teaching and Coherence in Learning*, *Cyprus, 2013,* edited by C. P. Constantinou, N. Papadouris and A. Hadjigeorgiou, Part 10, p. 180

[17] M. Taras, "Summative assessment: the missing link for formative assessment", Journal of Further and Higher Education, 33, 57 (2009)

[18] F. Dochy, M. Segers & D. Sluijsmans "The use of self-, peer and co-assessment in higher education: A review"; Studies in Higher Education, 24, 331 (1999)

[19] N. S. Rebello, "Comparing students' performance on research-based conceptual assessments and traditional classroom assessments" in *2011 PERC Proceedings*, *Omaha, 2011,* edited by N. S. Rebello, P. V. Engelhardt and C. Singh, p. 66

[20] L. Schuwirth and C. Van Der Vleuten, "Different written assessment methods: what can be said about their strengths and weaknesses?", Medical Education, 38, 974 (2004)

[21] B. R. Wilcox and S. J. Pollock , "Coupled multiple-response versus free-response conceptual assessment: An example from upper-division physics", Phys. Rev. ST Phys. Educ. Res. 10, 020124 (2014)

[22] A. Maries and C. Singh, "Exploring one aspect of pedagogical content knowledge of teaching assistants using the test of understanding graphs in kinematics" Phys. Rev. ST Phys. Educ. Res. 9, 020120 (2013)

[23] D. Hestenes, M. Wells, and G. Swackhammer, "Force Concept Inventory", Phys. Teach. 30, 141 (1992)

[24] L. Ding, R. Chabay, B. Sherwood and R. Beichner, "Evaluating an electricity and magnetism assessment tool: Brief electricity and magnetism assessment", Phys. Rev. ST Phys. Educ. Res. 2, 010105 (2006)

[25] A. D. Slepkov and R. C. Shiell, "Comparison of integrated testlet and constructed-response question formats", Phys. Rev. ST Phys. Educ. Res. 10, 020120 (2014)

[26] H. Kruglak, "Experimental Study of Multiple-Choice and Essay Tests", Am. J. Phys. 33, 1036 (1965)

[27] S. R. Wininger, "Using your tests to teach: Formative summative assessment" Teaching of Psychology, 32, 164 (2005)

[28] A. Pawl, R. E. Teodorescu and J. D. Peterson, "Assessing class-wide consistency and randomness in responses to true or false questions administered online", Phys. Rev. ST Phys. Educ. Res. 9, 020102 (2013)

[29] H. Yu, H. Li, "Group-based Formative Assessment: A Successful Way to Make Summative Assessment Effective", Theory and Practice in Language Studies, 4, 839 (2014)

[30] B. R. Wilcox and S. J. Pollock, "Upper-division student difficulties with the Dirac delta function", Phys. Rev. ST Phys. Educ. Res. 11, 010108 (2015)

[31] W. Fakcharoenphol and T. Stelzer, "Physics exam preparation: A comparison of three methods", Phys. Rev. ST Phys. Educ. Res. 10, 010108 (2014)

[32] K. E. Guest and D. S. Murphy, "In support of memory retention: A cooperative oral final exam" Education, 121, 350 (2000)

[33] M. Lusk and L. Conklin, "Collaborative testing to promote learning", Journal of Nursing Education; 42, 121 (2003)

[34] S. P. Rao, H. L. Collins, and S. E. DiCarlo, "Collaborative testing enhances student learning", Advances in Physiology Education, 26, 37 (2002)

[35] J. Ives, "Measuring the Learning from Two-Stage Collaborative Group Exams" in *2014 PERC Proceedings, Minneapolis, 2014,* edited by P. V. Engelhardt, A. D. Churukian, and D. L. Jones, p. 123

[36] R. N. Cortright, H. L. Collins, D. W. Rodenbaugh and S. E. DiCarlo, "Student retention of course content is improved by collaborative-group testing", Advances in Physiology Education, 27, 102 (2003)

[37] S. H. Kapitanoff, "Collaborative testing Cognitive and interpersonal processes related to enhanced test performance", Active Learning in Higher Education, 10, 56 (2009)

[38] B.T. Duane and M.E. Satre, "Utilizing constructivism learning theory in collaborative testing as a creative strategy to promote essential nursing skills", Nurse Education Today, 34, 31 (2014)

[39] B. H. Gilley, & B. Clarkston, "Collaborative testing: Evidence of learning in a controlled in-class study of undergraduate students", Journal of College Science Teaching, 43, 83 (2014)

[40] W. Jolliffe, *Cooperative Learning in the Classroom: Putting it into Practice*, (SAGE Publications Ltd, 2007), chapters: "Introduction: Cooperative Learning: What is it and Why Does it Matter?" , "Talk, Talk, Talk", "Putting Cooperative Learning into Practice"

[41] R. E. Slavin, *Cooperative Learning: Theory, research, and practice* (New Jersey: Prentice-Hall, 1990)

[42] S. Kagan, "The structural approach to cooperative learning", Educational Leadership, 47, 12 (1990)

[43] D. Halliday, R. Resnick, J. Walker, *Fundamentals of physics, part 3*, (John Wiley & Sons, Inc, 2001), p. 728, question 8

[44] R. R. Hake, "Interactive-engagement versus traditional methods: A sixthousand-student survey of mechanics test data for introductory physics courses," Am. J. Phys. 66, 64, (1998)

[45] F. W. Gery, "Does mathematics matter?", in *Research Papers in Economic Education*, edited by Arthur Welsh (Joint Council on Economic Education, New York, 1972), p.142

[46] D. Dallmer, "Collaborative test taking with adult learners", Adult Learning, 15, 4 (2004)

[47] J.V. Shindler, "Greater than the sum of parts? Examining the soundness of collaborative exams in teacher education courses", Innovative Higher Education, 28, 273 (2004)

[48] S.S. Sandahl, "Collaborative testing as a learning strategy in nursing education", Nursing Education Perspectives, 31, 142 (2010)

[49] P. G. Zimbardo, L. D. Butler and V. A. Wolfe, 'Cooperative College Examinations: More Gain, Less Pain When Students Share Information and Grades', Journal of Experimental Education 71, 101 (2003)

[50] J. C. Hendrix. "Cooperative Learning: Building a Democratic Community", The Clearing House, 69, 333 (1996)

[51] D. Bloom, "Collaborative test taking: benefits for learning and retention", College Teaching 57, 216 (2009)

[52] A. Haberyan, J. Barnett, "Collaborative testing and achievement: are two heads really better than one?", Journal of Instructional Psychology, 37, 32 (2010)

[53] M. Jensen, R. Moore and J. Hatch, "Cooperative Learning – Part I. Cooperative Quizzes", The American Biology Teacher, 64, 29 (2002)

[54] M. G. Simpkin, "An Experimental Study of the Effectiveness of Collaborative Testing in an Entry-Level Computer Programming Class", Journal of Information Systems, 16, 273 (2005)

[55] O. Dahlstrom, "Learning during a collaborative final exam", Educational Research and Evaluation, 18, 321, 2012